\documentclass[aps,prd,twocolumn,
comment, 
superscriptaddress,groupedaddress,
amsmath,amssymb,floatfix,nofootinbib]
{revtex4}  
\usepackage{float}
\usepackage{graphicx}
\usepackage{epstopdf}
\usepackage{dcolumn}
\DeclareGraphicsExtensions{.eps}
\usepackage{bm}
\usepackage[usenames,dvipsnames]{color}
\usepackage{slashed}
\usepackage{slashed}
\usepackage{url}
\usepackage{hhline}

\begin{document}

%

\let\a=\alpha      \let\b=\beta       \let\c=\chi        \let\d=\delta
\let\e=\varepsilon \let\f=\varphi     \let\g=\gamma      \let\h=\eta
\let\k=\kappa      \let\l=\lambda     \let\m=\mu
\let\o=\omega      \let\r=\varrho     \let\s=\sigma
\let\t=\tau        \let\th=\vartheta  \let\y=\upsilon    \let\x=\xi
\let\z=\zeta       \let\io=\iota      \let\vp=\varpi     \let\ro=\rho
\let\ph=\phi       \let\ep=\epsilon   \let\te=\theta
\let\n=\nu
\let\D=\Delta   \let\F=\Phi    \let\G=\Gamma  \let\L=\Lambda
\let\O=\Omega   \let\P=\Pi     \let\Ps=\Psi   \let\Si=\Sigma
\let\Th=\Theta  \let\X=\Xi     \let\Y=\Upsilon
%


\def\cA{{\cal A}}                \def\cB{{\cal B}}
\def\cC{{\cal C}}                \def\cD{{\cal D}}
\def\cE{{\cal E}}                \def\cF{{\cal F}}
\def\cG{{\cal G}}                \def\cH{{\cal H}}
\def\cI{{\cal I}}                \def\cJ{{\cal J}}
\def\cK{{\cal K}}                \def\cL{{\cal L}}
\def\cM{{\cal M}}                \def\cN{{\cal N}}
\def\cO{{\cal O}}                \def\cP{{\cal P}}
\def\cQ{{\cal Q}}                \def\cR{{\cal R}}
\def\cS{{\cal S}}                \def\cT{{\cal T}}
\def\cU{{\cal U}}                \def\cV{{\cal V}}
\def\cW{{\cal W}}                \def\cX{{\cal X}}
\def\cY{{\cal Y}}                \def\cZ{{\cal Z}}


\def\be{\begin{equation}}
\def\ee{\end{equation}}
\def\bea{\begin{eqnarray}}
\def\eea{\end{eqnarray}}
\def\bm{\begin{matrix}}
\def\em{\end{matrix}}
\def\bpm{\begin{pmatrix}}
    \def\epm{\end{pmatrix}}

{\newcommand{\lsim}{\mbox{\raisebox{-.6ex}{~$\stackrel{<}{\sim}$~}}}
{\newcommand{\gsim}{\mbox{\raisebox{-.6ex}{~$\stackrel{>}{\sim}$~}}}
\def\mpl{M_{\rm {Pl}}}
\def\gev{{\rm \,Ge\kern-0.125em V}}
\def\tev{{\rm \,Te\kern-0.125em V}}
\def\mev{{\rm \,Me\kern-0.125em V}}
\def\ev{\,{\rm eV}}

\title{\boldmath  Constraints on Scalar Leptoquark from Kaon Sector}
\author{Girish Kumar}
\email{girishk@prl.res.in}
\affiliation{Physical Research Laboratory, Navrangpura, Ahmedabad 380 009, India}
\affiliation{Indian Institute of Technology Gandhinagar, Chandkheda, Ahmedabad 382 424, India}

\begin{abstract}
Recently, several anomalies in flavor physics have been observed, and it was noticed that leptoquarks might account for the  deviations from the Standard Model. In this work, we examine the effects of new physics originating from a scalar leptoquark model on the kaon sector. The leptoquark we consider is a $\tev$-scale particle and within the reach of the LHC. We  use the existing experimental data on the several kaon processes including  $K^{0}-\bar{K}^{0}$ mixing, rare decays  $K^{+} \rightarrow \pi^{+}  \nu \bar{\nu}$, $ K_L \rightarrow \pi^0  \nu \bar{\nu}$, short-distance part of  $K_L\rightarrow \m^{+}\m^{-}$, and lepton-flavor-violating decay $K_L\rightarrow \m^{\pm} e^{\mp}$  to obtain useful constraints on the model.

\end{abstract}
\maketitle

\section{Introduction}
\label{sec1}
The discovery of the last missing piece, the Higgs boson, in the first run of  the LHC marks the completion of the Standard Model (SM) \cite{Aad:2012tfa, Chatrchyan:2012xdj}. Though the SM has been exceptionally successful in explaining the experimental data collected so far, there are many important questions which demand for physics beyond the SM (see, for example, \cite{Ricciardi:2015iwa}). Therefore, it is natural to consider the SM as the low-energy limit of a more general theory above the electroweak scale. The direct collider searches at the high-energy frontier (\tev-scale) have not found any new particle but, interestingly, there are some tantalizing hints towards new physics (NP) from high-precision low-energy experiments in the flavor sector. To be specific, in 2012,
 BaBar measured the ratios of branching fractions for the semitauonic decay of the $B$ meson, $\bar{B}\rightarrow D^{\ast}\tau \bar{\nu}$, 
  \be
\cR(D^{(\ast)})=\frac{{\rm BR}(\bar{B}\rightarrow D^{(\ast)}\tau \bar{\nu})}{{\rm BR}(\bar{B}\rightarrow D^{(\ast)} \ell \bar{\nu})},
\ee
with $\ell = e, \mu$, and reported  $2.0\sigma$ and $2.7\sigma$ excesses over the SM predictions in the measurements of $\cR(D)$ and $\cR(D^{\ast})$, respectively \cite{Lees:2012xj}. Very recently, these decays have been measured by BELLE \cite{Huschle:2015rga} and LHCb \cite{Aaij:2015yra}. These results are in agreement with each other and when combined together show a significant deviation from the SM. A summary of the measurements of $\cR(D^{(\ast)})$ done by different collaborations together with the SM predictions is given in Table \ref{tab1}. 
\begin{table}[]
 \label{tab1}
    \setlength{\tabcolsep}{2pt}
    \setlength\extrarowheight{13pt}   
         
      \begin{tabular}{ | c | c  c |}
         \hline
            & $\cR(D^{\ast})$ & $\cR(D)$  \\
          \hhline{|=|==|}

        LHCb  \cite{Aaij:2015yra}& 0.336 $\pm$ 0.027 $\pm$ 0.030  &  -  \\ 
        BaBar \cite{Lees:2012xj} & 0.332 $\pm$ 0.024 $\pm$ 0.018  & 0.440 $\pm$ 0.058 $\pm$ 0.042    \\ 
        BELLE \cite{Huschle:2015rga} &0.293 $\pm$ 0.038 $\pm$ 0.015  &0.375 $\pm$ 0.064 $\pm$ 0.026   \\ 
        SM  Pred.\cite{Fajfer:2012vx}& 0.252 $\pm$ 0.003 & 0.300 $\pm$ 0.010  \\
        \hline
      \end{tabular}
       \caption{Summary of experimental measurement for the ratios $R(D^{(\ast)})$ and the expectation in the SM. Here the first (second) errors are statistical (systematic).}  
   \end{table}

Another interesting  indirect  hint of  NP  comes from the data on $b\rightarrow s \m^{+}\m^{-}$ processes. The  LHCb  Collaboration  has  seen  a  $2.6\sigma$ departure from the SM prediction in the lepton flavor universality ratio $R_K={\rm BR}(\bar{B}\rightarrow \bar{K}\m^{+} \m^{-})/ {\rm BR}(\bar{B}\rightarrow \bar{K} e^{+} e^{-}) = 0.745^{+0.090}_{-0.074} \pm 0.036$ in the dilepton invariant mass bin $1\gev^2<q^2<6\gev^2$ \cite{Aaij:2014ora}. Though the individual branching fractions for $\bar{B}\rightarrow \bar{K}\m^{+}\m^{-}$ and $\bar{B}\rightarrow \bar{K} e^{+} e^{-}$ are marred with large hadronic uncertainties in the SM \cite{Bobeth:2007dw}, their ratio is a very clean observable and predicted to be $R_K = 1.0003\pm 0.0001$  \cite{Hiller:2003js, Bobeth:2007dw}. Also, the recent data on angular observables of four-body distribution in the process $(B \rightarrow K^{*} (\rightarrow K \pi)\ell^{+}\ell^{-}$ indicate some tension with the SM \cite{Aaij:2013qta, Aaij:2015oid}, particularly the deviation of $\sim3\sigma$  in two of the $q^2$ bins of angular observable $P_5^{'}$ \cite{DescotesGenon:2012zf} (also see Refs.~\cite{Matias:2012xw,Descotes-Genon:2013vna} for discussion on other angular observables for $B \rightarrow K^{*} \ell^{+}\ell^{-}$ decay). In the decay $B_s\rightarrow \phi\mu^{+}\mu^{-}$,  a deviation of $3.5\sigma$ significance with respect to the SM prediction has also been reported by LHCb \cite{Aaij:2015esa}. The model-independent global fits to the  data on $b\rightarrow s\m^{+}\m^{-}$ observables point towards a solution with NP that is favored over the SM by $\sim 4\sigma$ \cite{Descotes-Genon:2013wba,Altmannshofer:2015sma,Descotes-Genon:2015uva}.

Several NP scenarios have been proposed to explain these discrepancies. The excesses in $\cR(D^{(\ast)})$ have been explained in a generalized framework of 2HDM in Refs.~\cite{Celis:2012dk, Ko:2012sv, Crivellin:2012ye}, in the framework of the $R$-parity violating Minimal Supersymmetric Standard Model in Ref.~\cite{Deshpande:2012rr}, in the $E_6$-motivated Alternative Left-Right Symmetric Model (ALRSM) in Ref.~\cite{Hati:2015awg}, and using a model-independent approach \cite{Datta:2012qk,Bhattacharya:2015ida,Tanaka:2012nw,Freytsis:2015qca},  while in Refs.~\cite{Dorsner:2009cu,Fajfer:2012jt, Sakaki:2013bfa, Dorsner:2013tla} the excesses in $\cR(D^{(\ast)})$ have been addressed in the context of leptoquark models. The possible explanation for the observed anomalies in $b\rightarrow s\m^{+}\m^{-}$ processes preferably demands a negative contribution to the Wilson coefficient of semileptonic operator $(\bar{s}b)_{\rm V-A}(\bar{\m}\g_\a\m)$ \cite{Blake:2015tda, Altmannshofer:2015sma}. Several NP models, generally involving $Z^{'}$ vector bosons \cite{Gauld:2013qja, Glashow:2014iga, Crivellin:2015lwa, Altmannshofer:2014cfa, Buras:2013qja, Crivellin:2015mga, Chiang:2016qov,Boucenna:2016wpr} or leptoquarks \cite{Gripaios:2014tna, Becirevic:2015asa, Alonso:2015sja, Calibbi:2015kma, Bauer:2015knc, Barbieri:2015yvd, Varzielas:2015iva, Fajfer:2015ycq, Hiller:2014yaa}, are able to produce such operators with the required effects to explain the present data.

 In view of this, we are motivated to study a $\tev$-scale leptoquark model and analyze NP effects on the kaon sector. It is known that the studies of kaon decays have played a vital role in retrieving information on the flavor structure of the SM. In particular, neutral kaon mixing and the rare decays of the kaon have been analyzed  in various extensions of the SM and are known to provide some of the most stringent constraints on NP \cite{Buras:2013ooa, Blanke:2013goa,Buras:1997ij, Buras:2012ts, Buras:2004qb, Blanke:2009am, Blanke:2008yr, Bauer:2009cf, Buras:2012dp, Buras:2014yna, Queiroz:2014pra, Lee:2015qra,Mescia:2012fg}. The NP model we consider in this paper is a simple extension of the SM by a single scalar leptoquark. The leptoquark $\phi$ with mass $M_\phi$ has $(SU(3),\,SU(2))_{U(1)}$ quantum numbers $(3,1)_{-1/3}$. This model is interesting considering that it has all the necessary ingredients accommodating semileptonic $b\rightarrow c$ and $b\rightarrow s$ decays to explain the anomalies in the LFU ratios discussed above \cite{Bauer:2015knc, Hiller:2014yaa}. To this end, we must mention that, along with anomalies observed in the flavor sector, the leptoquark model under study is also capable of explaining the new diphoton  excess recently reported by  the ATLAS and CMS collaborations in their analysis of  $\sqrt{s} = 13 \tev$ $pp$ collision \cite{diphoton excess ref}.  
 
 Following the conventions of Ref.~\cite{Bauer:2015knc}, the  Lagrangian governing the leptoquark interaction with first-family fermions is given by
\bea
\label{0.1}
\cL^{(\phi)} \ni \l^L_{ue}\bar{u}^c_L e_L \phi^{*} + \l^R_{ue}\bar{u}^c_R e_R \phi^{*} - \l^L_{d\nu}\bar{d}^c_L \nu_L \phi^{*} + h.c.,\nonumber\\
\eea
where L/R are the left/right projection operators $(1\mp \g_5)/2$. The couplings $\l$'s are family dependent, and $u^c$=$C\bar{u}^T$ are the charge-conjugated spinors. Similar interaction terms for the second and third families can also be written down. In this model, $B\rightarrow D^{(\ast)} \t\bar{\nu}$ proceeds at tree level through the exchange of leptoquark $(\phi)$. Integrating out the heavy particles gives rise to low-energy dimension-six effective operators, which can produce the required effects to satisfy the experimental data. In Ref.~\cite{Bauer:2015knc}, it was shown that  with $O(1)$ left-handed and relatively suppressed right-handed couplings one can explain the observed excesses in the rate of $B\rightarrow D^{(\ast)}\t\bar{\n}$ decays. The authors of Ref.~\cite{Bauer:2015knc} were also able to simultaneously explain the observed anomalies in $R_K$ with large $(\sim O(1))$ left-handed couplings for a $\tev$-scale leptoquark. In this model, such large couplings are possible because the leading contribution to $\bar{B}\rightarrow \bar{K}\m^{+}\m^{-}$ comes from one-loop diagrams and therefore additional GIM and CKM suppression compensates for the ``largeness" of the couplings. This is in contrast to NP models \cite{Hiller:2014yaa, Sahoo:2015wya, Becirevic:2015asa} in which $R_K$ arises at tree level, which renders the couplings very small in order to have leptoquarks within the reach of the LHC.  Apart from the  B system, this model has also been explored in the context of flavor changing neutral current (FCNC) decays of the D meson. In Refs.~\cite{Fajfer:2008tm, deBoer:2015boa, Fajfer:2015mia}, the  impact of scalar (as well as vector) leptoquarks  on the FCNC processes $D^{0}\rightarrow \m^{+} \m^{-}$  and $D^{+}\rightarrow \pi^{+} \m^{+} \m^{-}$ have been studied, and using the existing experimental results,  strong bounds on the leptoquark coupling  have been derived. However, to the best of our knowledge, the effects of new physics on the kaon sector have not been investigated before in the scalar leptoquark $(3,1)_{-1/3}$ model. We start by writing the effective Hamiltonian relevant for each case and discuss the effective operators  and  corresponding coupling strengths (Wilson coefficients) generated in the model. The explicit expressions of new contributions in terms of parameters of the model are derived.  We then discuss NP affecting the various kaon processes such as  $K^{+} \rightarrow \pi^{+}  \nu \bar{\nu}$, $ K_L \rightarrow \pi^0  \nu \bar{\nu}$, $K_L\rightarrow \m^{+}\m^{-}$, and LFV decay $K_L\rightarrow \m^{\pm} e^{\mp}$. Using the existing experimental information on these processes, the constraints on the leptoquark couplings are obtained. 

The rest of the article is organized in the following way. In section \ref{sec2}, we study the $K^0-\bar{K}^0$ mixing in this model and obtain constraints on the couplings. In sections \ref{sec3} and \ref{sec4}, we constrain the parameter space using information on ${\rm BR}(K^{+}\rightarrow \pi^{+}\nu\bar{\nu})$ and CP-violating ${\rm BR}(K_L\rightarrow \pi^0\nu\bar{\nu})$, respectively. In section \ref{sec5}, we discuss the new contribution to the short-distance part of rare decay $K_L\rightarrow\m^{+} \m^{-}$ in this model and obtain constraints on the generation-diagonal leptoquark couplings using the bounds on ${\rm BR}(K_L\rightarrow \m^{+}\m^{-})_{\rm SD}$. In section \ref{sec6}, we discuss  the LFV process $K_L\rightarrow \m^\mp e^\pm$ and constrain the off-diagonal couplings of the leptoquark contributing to NP Wilson coefficients. Finally, we summarize our results in the last section.

\section{Constraints From $K^0-\bar{K^0}$ Mixing}
\label{sec2}

The phenomenon of meson-antimeson oscillation, being a FCNC process, is very sensitive to heavy particles propagating in the mixing amplitude, and therefore it provides a powerful tool to test the SM and a window to observe NP. In this section, we focus on the mixing of the neutral kaon meson. The experimental measurement of the $K^0-\bar{K}^0$ mass difference $\Delta m_K$ and of CP-violating parameter $\epsilon_K$ has been instrumental in not only constraining the parameters of the unitarity triangle  but also providing stringent constraints on NP. The theoretical calculations for  $K^0-\bar{K}^0$ mixing are done in the framework of effective field theories (EFT), which allow one to separate long- and short-distance contributions. The leading contribution to $K^0-\bar{K}^0$ oscillations in the SM comes from the so-called box diagrams generated through internal line exchange of the $W$ boson and up-type quark pair. The effective SM Hamiltonian for $|\Delta S| = 2$ resulting from the evaluation of box diagrams is written as \cite{Antonelli:2009ws, Buchalla:1995vs}
\bea \label{2.1}
\cH_{\rm eff}^{|\Delta S|=2} &=& \frac{G_F^2 M_W^2}{4 \pi^2} \left(\l_c^2 \eta_{cc} S_0(x_c) + \l_t^2 \eta_{tt} S_0(x_t)\right.\nonumber \\
 && \hskip0.5cm + \left. 2\l_t \l_c \eta_{ct} S_0(x_c, x_t)\right) K(\mu) Q_s(\mu),
\eea
where $G_F$ is the Fermi constant and $\l_i = V_{is}^{*} V_{id}$ contains CKM matrix elements. $Q_s(\mu)$ is a dimension-six, four-fermion local operator $(\bar{s}\g_\mu L d)(\bar{s}\g^\mu L d)$,  and $K(\mu)$ is the relevant short-distance factor which makes product $K(\mu)Q_s(\mu)$ independent of $\mu$. The Inami-Lim functions $S_0(x)$ and $S_0(x_i, x_j)$  \cite{Inami:1980fz} contain contributions of loop diagrams and are given by \cite{Branco:1999fs}
\bea\label{2.2}
S(x_c, x_t) &=& x_c x_t \left[-\frac{3}{4(1-x_c)(1-x_t)} \right.\nonumber\\
&& +  \frac{{\rm Ln}\, x_t}{(x_t-x_c)(1-x_t)^2}\left(1-2x_t +\frac{x_t^2}{4}\right)\nonumber\\
&& + \left. \frac{{\rm Ln}\, x_c}{(x_c-x_t)(1-x_c)^2}\left(1-2x_c +\frac{x_c^2}{4}\right)\right],\nonumber\\
\eea
and the function $S_0(x)$ is the limit when $y\rightarrow x$ of $S_0(x,y)$, while $\eta_i$ in Eq.~\eqref{2.1} are the short-distance QCD correction factors $\eta_{cc} = 1.87$, $\eta_{tt} = 0.57$, and $\eta_{ct} = 0.49$  \cite{Buras:1990fn, Brod:2011ty, Brod:2010mj}. The hadronic matrix element $\langle\bar{K}^0\lvert Q_s\rvert K^0\rangle$ is parametrized in terms of decay constant $f_K$ and kaon bag parameter $B_K$ in the following way:
\be
\label{2.3}
B_K = \frac{3}{2} K(\mu) \frac{\langle\bar{K}^0\lvert Q_s\rvert K^0\rangle}{f_K^2 m_K^2}.
\ee

The contribution of NP to $|\D S|=2$ transition can be parametrized as the ratio of the full amplitude to the SM one as follows \cite{Bona:2007vi}:
\be
\label{2.4}
C_{\D m_K} = \frac{{\rm Re}\langle\ K\lvert H_{\rm eff}^{\rm Full}\rvert \bar{K}\rangle}{{\rm Re}\langle K\lvert H_{\rm eff}^{\rm SM}\rvert \bar{K}\rangle},\,\,\,\, C_{\e_K} = \frac{{\rm Im}\langle K\lvert H_{\rm eff}^{\rm Full}\rvert \bar{K}\rangle}{{\rm Im}\langle K\lvert H_{\rm eff}^{\rm SM}\rvert \bar{K}\rangle},
\ee
In the SM, $C_{\D m_K}$ and $C_{\e_K}$ are unity. The effective Hamiltonian $\langle\bar{K}^0\lvert H_{\rm eff}\rvert K^0\rangle$ can be related to the off-diagonal element $M_{12}$ through the relation \footnote{The observables mass difference $\D m_K$ and CP-violating parameter $\e_K$ are related to off-diagonal element $M_{12}$ through the following relations: $\D m_K = 2 [{\rm Re}(M_{12})_{SM} + {\rm Re}(M_{12})_{NP}]$ and $\e_K = \frac{k_\e \exp^{i \phi_\e}}{\sqrt{2}(\D m_K)_{\rm exp}} [{\rm Im}(M_{12})_{SM} + {\rm Im}(M_{12})_{NP}]$, where $\phi_\e \simeq 43^{o}$ and $k_\e \simeq 0.94$ \cite{Buras:2008nn, Buras:2010pza, Ligeti:2016qpi}.}
\be
\label{2.5}
\langle\bar{K}^0\lvert H_{\rm eff}^{\rm Full}\rvert K^0\rangle = 2 m_K M_{12}^{\ast},
\ee
with $M_{12} = (M_{12})_{SM} + (M_{12})_{NP}$. In the SM, the theoretical expression of $(M_{12})_{SM}$ reads \cite{Buras:2013ooa}
\be
\label{2.6}
(M_{12})_{SM} = \frac{G_F^2}{12 \pi^2} f_K^2 B_K m_K M_W^2 F^{*}(\l_c, \l_t, x_c, x_t),
\ee
where the function $F(\l_c, \l_t, x_c, x_t)$ stands for
\bea\label{2.7}
F(\l_c, \l_t, x_c, x_t) &=& \l_c^2 \eta_{cc} S_0(x_c) + \l_t^2 \eta_{tt} S_0(x_t) \nonumber\\
&& \hskip0.5cm + \,\,\, 2\l_t \l_c \eta_{ct} S_0(x_c, x_t),
\eea
with $x_i \!=\!\! m_i^2/M_W^2$. 

In the $(3,1)_{-1/3}$ leptoquark model, the internal line exchange of the neutrino-leptoquark pair induces new Feynman diagrams, which contributes to  $K^0-\bar{K}^0$ mixing. The diagrams are shown in Fig~\ref{fig1}.
 \begin{figure}[ht!]
 	\centering
 	\hspace{0.02cm}
 	\hbox{\hspace{0.02cm}
 		\hbox{\includegraphics[scale=0.39]{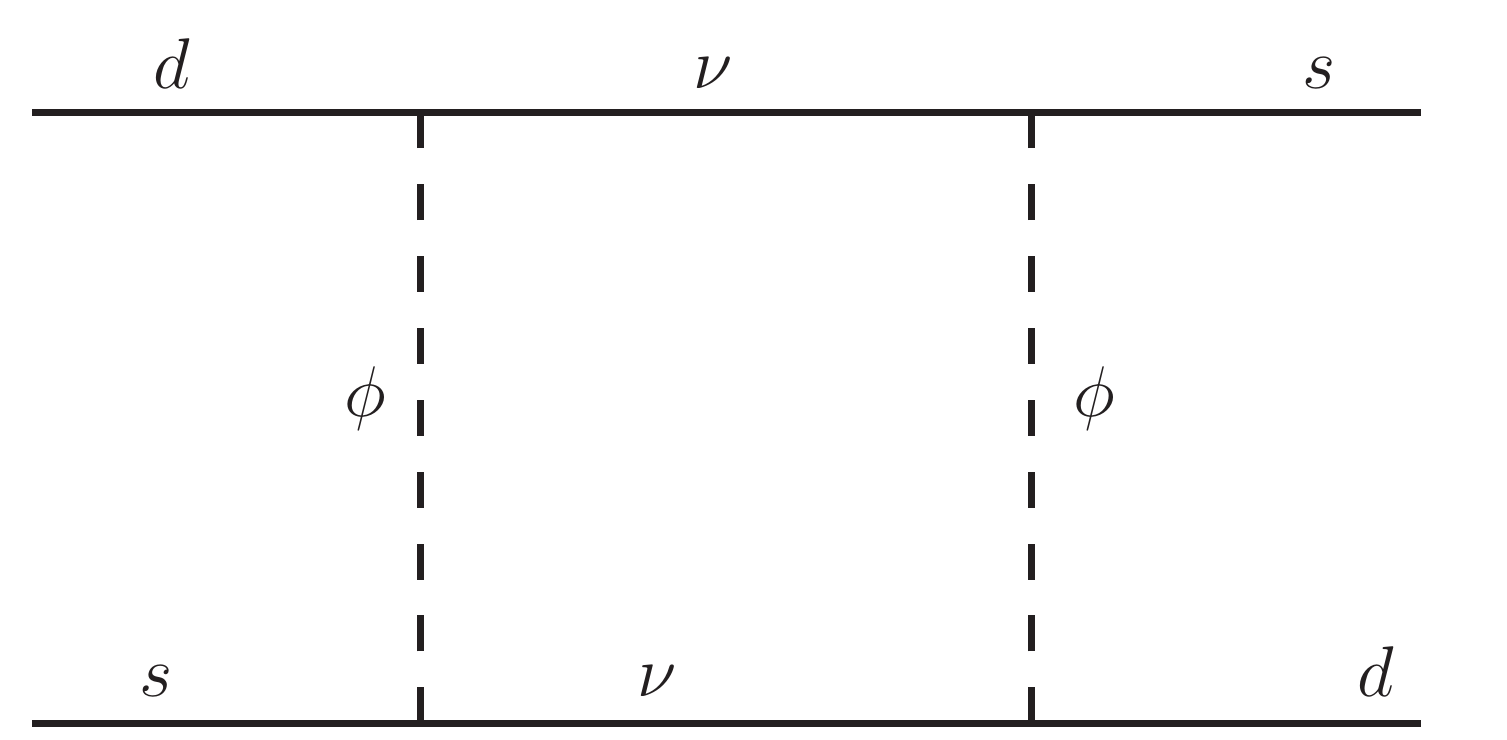}}
 	}
 	
 	\caption{ New contribution to $K-\bar{K}$ mixing induced by the scalar leptoquark $(\phi)$.
 	}
 	\label{fig1}
 \end{figure}
  The new effects modify the observables $C_{\D m_K}$ and $C_{\e_K}$, and in the approximation $M_\phi^2 \gg m^2_{t, W}$, their expressions are given by

  \bea
  \label{2.8}
  C_{\D m_K} &=& 1 + \frac{1}{g_2^4} \frac{M_W^2}{M_\phi^2}\frac{\eta_{tt}}{{\rm Re}(F^{*})}{\rm Re}\left(\xi_{ds}\right)^2,\\
  C_{\e_K} &=& 1 + \frac{1}{g_2^4} \frac{M_W^2}{M_\phi^2}\frac{\eta_{tt}}{{\rm Im}(F^{*})}{\rm Im}\left(\xi_{ds}\right)^2,
  \eea
where we have used notation $F$ for $F(\l_c, \l_t, x_c, x_t)$ for simplicity. $g_2$ is the SU(2) gauge coupling and we define
\be
\label{2.9.1}
\xi_{ds} \equiv (\l^L\l^{L \dagger})_{ds} = \sum_i \l^L_{d\n_i}\l^{L\ast}_{s\n_i}.
\ee

 Solving Eqs.~(10) and (11) for real and imaginary parts of $\xi_{ds}$ in terms of the experimental observables $C_{\D m_K}$ and $C_{\e_K}$, we obtain the following expressions:
\bea
\label{2.10}
\left({\rm Re}\,\xi_{ds}\right)^2 &=&  \left(\frac{g_2^4}{2}\frac{M_\phi^2}{M_W^2}\right)\Biggl(\frac{{\rm Re}(F^{*})}{\eta_{tt}}\Bigl(-1 + C_{\D m_K}\Bigr)\Biggr)\nonumber\\
&& \times  \Biggl(1 + \sqrt{1 + \left(\frac{{\rm Im} F^{*}}{{\rm Re} F^{*}}\cdot\frac{C_{\e_K} -1}{C_{\D m_K}-1}\right)^2}\,\, \Biggr),\nonumber\\
\\
\left({\rm Im}\,\xi_{ds}\right)^2 &=& \left(\frac{g_2^4}{2}\frac{M_\phi^2}{M_W^2}\right)\Biggl(\frac{{\rm Re}(F^{*})}{\eta_{tt}}\Bigl(-1 + C_{\D m_K}\Bigr)\Biggr)\nonumber\\
&& \times  \Biggl(-1 + \sqrt{1 + \left(\frac{{\rm Im} F^{*}}{{\rm Re} F^{*}}\cdot\frac{C_{\e_K} -1}{C_{\D m_K}-1}\right)^2}\,\, \Biggr).\nonumber\\
\eea

To constrain the leptoquark couplings $\rm{Re}\,\xi_{ds}$ and  $\rm{Im}\,\xi_{ds}$, we use the latest global fit results  provided by the UTfit collaboration, and to be conservative evaluate the constraints at the 2$\sigma$ level:  $C_{\D m_K} = 1.10 \pm 0.44$ and  $C_{\e_K} = 1.05 \pm 0.32$ \cite{Bona:2007vi}.  Here, to account for the significant uncertainties from poorly  known long-distance effects \cite{Buras:2014maa}, we allow for a $\pm 40\%$ uncertainty in the case of $\Delta M_K$. For ${\rm Re}\,\xi_{ds}$ and ${\rm Im}\,\xi_{ds}$, we obtain the following upper bounds:
\bea
\label{2.11}
\left({\rm Re}\,\xi_{ds}\right)^2 \le 6.0\times 10^{-4} \left(\frac{M_\phi}{1000 \gev}\right)^2,\\ \nonumber\\
\left({\rm Im}\,\xi_{ds}\right)^2 \le 3.8\times 10^{-4} \left(\frac{M_\phi}{1000 \gev}\right)^2.
\eea

As discussed in the next section, we find that a more constraining bound on the product of the couplings ${\rm Re}(\xi_{ds})$ and ${\rm Im}(\xi_{ds})$ can be obtained from theoretically rather clean rare processes $K^{+} \rightarrow \pi^{+} \nu \bar{\nu}$ and $K_L \rightarrow \pi^{0} \nu \bar{\nu}$ as compared to $K-\bar{K}$ mixing.

\section{{\boldmath Constraints from rare decay $K^{+} \rightarrow \pi^{+} \nu \bar{\nu}$}}
\label{sec3}
The charged and neutral $K\rightarrow \pi \nu \bar{\nu}$ are in many ways interesting FCNC processes and considered as {\it golden} modes. Both the decays can play an important role in indirect searches for NP because these decays are theoretically very clean and their branching ratio can be computed with an exceptionally high level of precision (for a review, see Ref.~\cite{Buras:2004uu}). In the SM, these decays are dominated by Z-penguin and box diagrams, which exhibit hard, powerlike GIM suppression as compared to logarithmic GIM suppression generally seen in other loop-induced meson decays. At the leading order, both modes are induced by a single dimension-six local operator $(\bar{s}d)_{\rm V-A} (\bar{\nu}\nu)_{\rm V-A}$. The hadronic matrix element of this operator can be measured precisely in $K^{+}\rightarrow \pi^0 e^{+} \nu$ decays, including isospin breaking corrections \cite{Mescia:2007kn, Isidori:2005xm}. The principal contribution to the error in theoretical predictions originates from the uncertainties on the current values of $\l_t$ and $m_c$. The long-distance effects are rather suppressed and have been found to be small \cite{Buchalla:1998ux, Geng:1996kd, Buchalla:1997kz}. 

In the SM, the effective Hamiltonian for $K \rightarrow \pi \nu \bar{\nu}$ decays is written as \cite{Buchalla:1993wq}
\bea
\label{3.1}
\cH_{\rm eff}^{\rm SM} &=& \frac{G_F}{\sqrt{2}}\frac{2\a}{\pi \sin^2\theta_W} \sum_{\ell = e,\mu,\tau} \left(\l_c X_{\rm NNL}^\ell + \l_t X(x_t)\right) \nonumber\\
&& \hskip2cm\times (\bar{s}_L\g_\mu d_L) ( \bar{\nu}_{\ell L}\g^\mu\nu_{\ell L}),
\eea
The index $\ell = e, \mu, \tau$ denotes the lepton flavor. The short-distance function $X(x_t)$ corresponds to the loop-function containing top contribution and is given by
\bea
\label{3.2}
X(x_t) = \eta_X\cdot \frac{x_t}{8} \left[\frac{x_t +2}{x_t-1}+\frac{3x_t-6}{(x_t-1)^2}{\rm Ln}\,x_t\right],
\eea
where the factor $\eta_X$ includes the NLO correction and is close to unity ($\eta_X = 0.995$), while the remaining part describes the contribution of top quark without QCD correction. The NLO QCD corrections have been computed in Refs.~\cite{Buchalla:1993bv, Misiak:1999yg, Buchalla:1998ba}, while two-loop electroweak corrections have been studied in Ref.~\cite{Brod:2010hi}. The loop-function $X_{\rm NNL}$ summarizes the contribution from the charm quark and can be written as \cite{Buras:1997ij}
\bea
\label{3.3}
X_{NNL} = \frac{2}{3} X_{NNL}^e + \frac{1}{3} X_{NNL}^\tau \equiv \l^4 P_c^{SD}(X),
\eea
where $\l = |V_{us}|$. The NLO results for the function $X_{\rm NNL}$ can be found in Refs.~\cite{Buchalla:1998ba, Buchalla:1993wq}, while NNLO calculations are done in Refs.~\cite{Buras:2005gr, Buras:2006gb}. \newline

In the considered model, leptoquark $\phi$ mediates $K^{+}\rightarrow \pi^{+} \nu \bar{\nu}$ at tree level. The corresponding Feynman diagram is shown in Fig~\ref{fig2}. Integrating out the heavy degrees of freedom, we obtain the following NP effective Hamiltonian relevant for $K^{+}\rightarrow \pi^{+} \nu \bar{\nu}$ decay:
 \begin{figure}[ht!]
 	\centering
 	\hspace{0.02cm}
 	\hbox{\hspace{0.02cm}
 		\hbox{\includegraphics[scale=0.50]{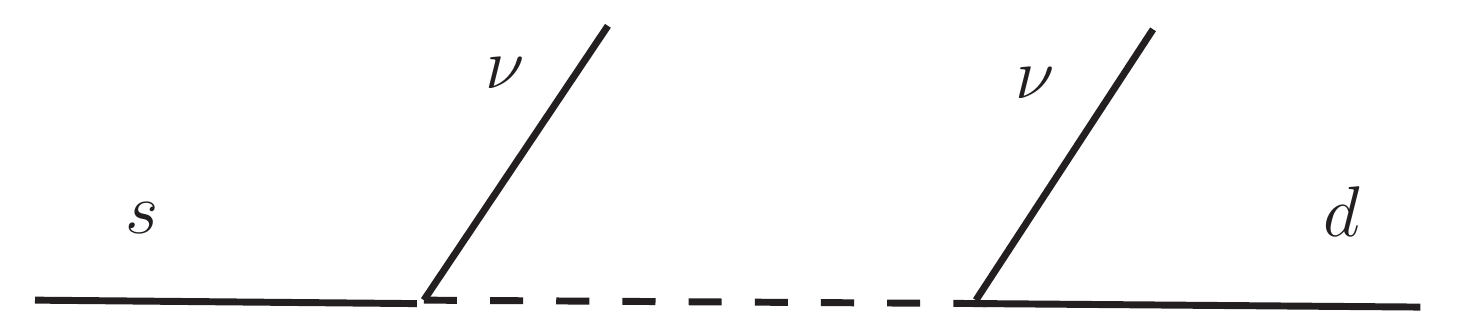}}
 	}
 	
 	\caption{ Feynman diagrams for the decay $K\rightarrow \pi \nu \bar{\nu}$ induced by the exchange of  scalar leptoquark $\phi$.
 	}
 	\label{fig2}
 \end{figure}

\bea
\label{3.4}
\cH_{\rm eff}^{\rm NP} = -\frac{\l^{L\ast}_{s\nu_\ell}\l^{L}_{d\nu_\ell}}{2M_\phi^2} (\bar{s}\g_\mu L d)(\bar{s}\g^\mu L d).
\eea
The new contribution alters the SM branching ratio of  $K^{+}\rightarrow \pi^{+} \nu \bar{\nu}$  \cite{Buras:2015qea} as
\bea
\label{3.5}
{\rm BR} (K^{+}\rightarrow \pi^{+} \nu \bar{\nu}) &=& \k_{+}(1+\D_{\rm EM})\left[ \left(\frac{{\rm Im}\,\l_t}{\l^5}X_{\rm new}\right)^2\right.\nonumber\\
 && + \left.\left(\frac{{\rm Re}\,\l_c}{\l}P_c(X)+\frac{{\rm Re}\,\l_t}{\l^5}X_{\rm new}\right)^2\right],\nonumber\\
\eea
where $\k_{+}$ contains relevant hadronic matrix elements extracted from  the decay rate of $K^{+} \rightarrow \pi^{0}e^{+}\nu$ along with isospin-breaking correction factor. The explicit form of $\k_{+}$ can be found in Ref.~\cite{Buras:1998raa}. $\D_{\rm EM}$ describes the electromagnetic radiative correction from photon exchanges and amounts to -0.3\%. The charm contribution $P_c(X)$ includes the short-distance part $P_c^{\rm SD}(X)$ plus the long-distance contribution $\d P_c$ (calculated in Ref.~\cite{Isidori:2005xm}). We use $P_c(X) = 0.404$ given in Ref.~\cite{Buras:2015qea}. The function $X_{\rm new}$ contains a new short-distance contribution from the leptoquark-mediated diagram and modifies the SM contribution through
 \begin{figure}[ht!]
 	\centering
 	
 	\hbox{\!\!
 		\hbox{\includegraphics[scale=0.36]{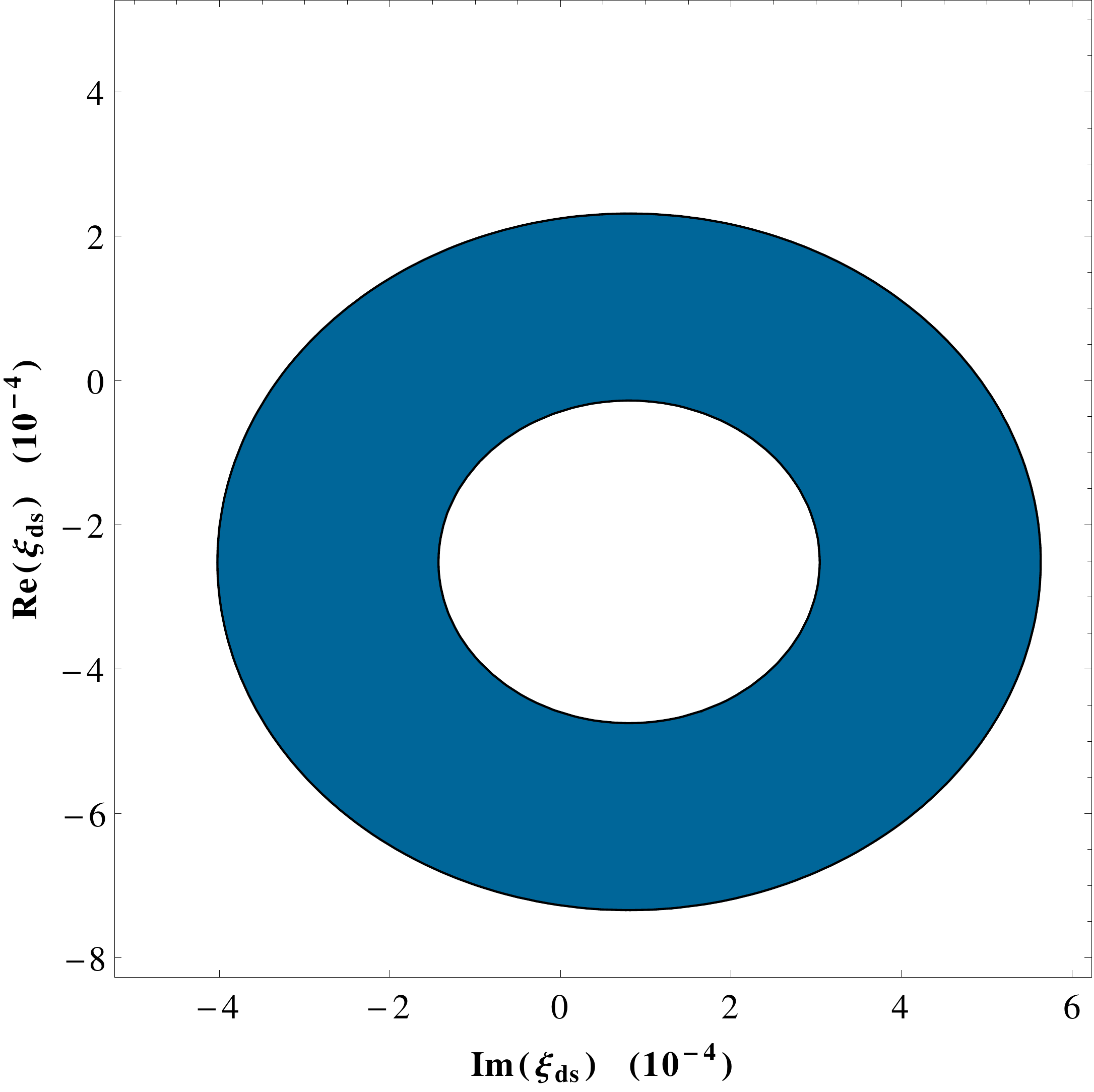}}
 	}
 	
 	\caption{ The constraints on ${\rm Re}(\xi_{ds})-{\rm Im}(\xi_{ds})$ parameter space from the measured value of ${\rm BR }(K^{+}\rightarrow\pi^{+}\nu\bar{\nu})$. The blue colored region shows experimentally allowed values at the $1\sigma$ level.
 	}
 	\label{fig3}
 \end{figure}

\bea
\label{3.6}
X_{new} = X(x_t) + \frac{X_{\phi}}{\l_t},
\eea
where $X(x_t)$ is the top contribution in the SM already defined in Eq. \eqref{3.2} and $X_\phi$ is the contribution due to leptoquark exchange. In terms of the model parameters, $X_\phi$ is given by

\bea
\label{3.7}
X_\phi = -\frac{\sqrt{2}}{4 G_F}\frac{\pi \sin^2\theta_W}{\a}\frac{\xi_{ds}}{M_\phi^2},
\eea
where $\a(M_Z) = 1/127.9$ is the electromagnetic coupling constant, and   $\sin^2\theta_W = 0.23$ is the weak mixing angle.  Using the experimental value of the branching ratio from the Particle Data Group,  ${\rm BR}(K^{+}\rightarrow\pi^{+}\nu\bar{\nu}) = (1.7\pm1.1)\times10^{-10}$ \cite{Agashe:2014kda}, we obtain the constraint on ${\rm Re}\,\xi_{ds}$ and ${\rm Im}\,\xi_{ds}$, shown in Fig~\ref{fig3}. A most conservative bound on individual couplings ${\rm Re}\,\xi_{ds}$ and ${\rm Im}\,\xi_{ds}$ can be obtained by taking  only one set to be nonzero at a time. We find that for a leptoquark of $1\tev$  mass the constraints are given by $-7.2\times 10^{-4} < {\rm Re}\,\xi_{ds} < 2.2 \times 10^{-4}$ and $-3.3\times 10^{-4} < {\rm Im}\,\xi_{ds} < 4.9 \times 10^{-4}$. As pointed out before, these bounds rule out a large parameter space allowed from $K^0-\bar{K}^0$ mixing. The coupling ${\rm Im}\,\xi_{ds}$ can also be probed independently through the decay $K_L \rightarrow \pi^{0} \nu \bar{\nu}$, which is the subject of our next section.

 \section{{\boldmath Constraints from  $K_L \rightarrow \pi^{0} \nu \bar{\nu}$}}
 \label{sec4}
The neutral decay mode $K_L \rightarrow \pi^{0} \nu \bar{\nu}$ is CP-violating. In contrast to the decay rate of  $K^{+}\rightarrow\pi^{+}\nu\bar{\nu}$ which depends on the real and imaginary parts of $\l_t$, with a small contribution from the real part of $\l_c$, the rate of $K_L \rightarrow \pi^{0} \nu \bar{\nu}$ depends only on Im$\l_t$. Because of the absence of the charm contribution, the prediction for ${\rm BR}(K_L \rightarrow \pi^{0} \nu \bar{\nu})$ is theoretically cleaner. The principal sources of error are the uncertainties on Im$\l_t$ and $m_t$. In the SM, the branching ratio is given by \cite{Buras:2004uu}
\bea
\label{3.8}
{\rm BR} (K_L\rightarrow \pi \nu \bar{\nu}) &=& \k_{L} \left(\frac{{\rm Im}\l_t}{\l^5}X(x_t)\right)^2,
\eea
with \cite{Buras:2015qea}
\bea
\label{3.9}
\k_L = 2.231 \times 10^{-10}\left(\frac{\l}{0.225}\right)^8.
\eea

The exchange of leptoquark $\phi$ induces new contribution to the rate which can be accommodated in the expression of branching ratio  by replacing $X(x_t)$ with $X_{\rm new}$ given in Eq.~\eqref{3.6}. Experimentally, only a upper bound on the branching ratio is available: ${\rm BR}(K_L\rightarrow\pi^{0}\nu\bar{\nu}) < 2.8\times10^{-8}$ at 90\% C.L. \cite{Agashe:2014kda}. In Fig~\ref{fig4}, we plot the dependence of  $K_L\rightarrow \pi \nu \bar{\nu}$ branching ratio on the imaginary part of the effective couplings $\xi_{ds}$. Numerically, the constraints are given by

\begin{figure}[ht!]
 	\centering
 	
 	\hbox{\!\!
 		\hbox{\includegraphics[scale=0.35]{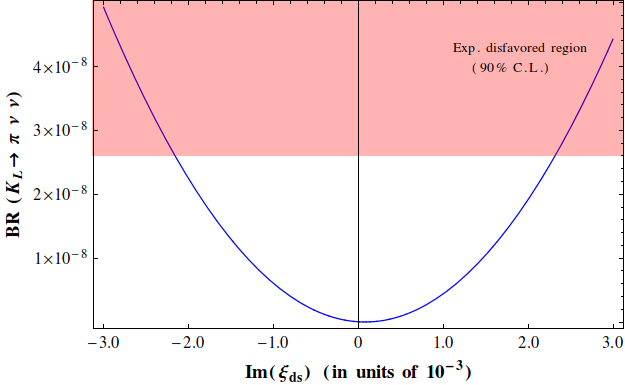}}
 	}
 	
 	\caption{ The dependence of ${\rm BR}(K_L\rightarrow\pi^{0}\nu\bar{\nu})$ on ${\rm Im}\,\xi_{ds}$. The red shaded region is currently disfavored by the experimental data at 90\% C.L. 
 	}
 	\label{fig4}
 	 	
 \end{figure}
 
 	\bea
  	\label{3.10}
  -0.0021  <	\frac{{\rm Im}\,\xi_{ds}}{\left(\frac{M_\phi}{1000 \gev}\right)^2}  < 0.0023,
  	\eea

Since the decay has not been observed so far and the present experimental limits are 3 orders of magnitude above the SM predictions \cite{Buras:2015qea}, we find that constraints from  $K_L \rightarrow \pi^0\nu\bar{\nu}$ are weaker compared to those obtained in the case of $K^{+} \rightarrow \pi^{+}\nu\bar{\nu}$. 

\section{{\boldmath Constraints from $K_L \rightarrow \mu^{+} \mu^{-}$}}
\label{sec5}
The decay $K_L\rightarrow \mu^{+}\mu^{-}$ is sensitive to much of the same short-distance physics (i.e., $\l_t$ and $m_t$) as $K\rightarrow\pi \nu\bar{\nu}$ and therefore provides complementary information on the structure of FCNC $\lvert\D S\rvert=1$ transitions. This is important because experimentally a much more precise measurement compared to $K\rightarrow\pi \nu\bar{\nu}$ is available: ${\rm BR}(K_L\rightarrow\mu\mu) = (6.84\pm0.11)\times10^{-9}$ \cite{Agashe:2014kda}. However, the theoretical situation is far more complex (for a review, see Refs.~\cite{Ritchie:1993ua, Cirigliano:2011ny}). The amplitude for  $K_L\rightarrow\mu^{+}\mu^{-}$ can be decomposed into a dispersive (real) and an absorptive (imaginary) part. The dominant contribution to the absorptive part (as well as to total decay rate ($K_L\rightarrow\mu^{+}\mu^{-}$)) comes from the real two-photon intermediate state. The dispersive amplitude is the sum of the so-called long-distance and the short-distance contributions. Only the short-distance (SD) part can be reliably calculated. The most recent estimates of the SD part from data give ${\rm BR}(K_L\rightarrow\mu^{+}\mu^{-})_{\rm SD} \le 2.5\times10^{-9}$ \cite{Isidori:2003ts}. The effective Hamiltonian relevant for the decay $K_L\rightarrow\mu^{+}\mu^{-}$ is given by \cite{Buchalla:1993wq}

\bea
\label{4.1}
\cH_{\rm eff}(K_L \rightarrow \mu^{+} \mu^{-}) &=& \frac{G_F}{\sqrt{2}}\frac{\a}{2\pi\sin^2\!\theta_W}(\l_c Y_{NL}+\l_tY(x_t))\nonumber\\
&& \hskip0.8cm\times (\bar{s}\g^\mu(1-\g_5)d)(\bar{\mu}\g_\mu\g_5\mu),\nonumber\\
&=& \frac{G_F}{\sqrt{2}}V^{*}_{us}V_{ud}\,\,\D^K_{SM}\nonumber\\
&& \hskip0.5cm\times (\bar{s}\g^\mu(1-\g_5)d)(\bar{\mu}\g_\mu\g_5\mu),\nonumber\\
\eea
where $\D^K_{SM}$ describes the Wilson coefficient (WC) of the effective local operator  $(\bar{s}d)_{\rm V-A}(\bar{\mu}\g_\mu\g_5\mu)$ and is given as
\bea
\label{4.2}
\D^K_{SM}=\frac{\a(\l_c Y_{NL}+\l_tY(x_t))}{2\pi\sin^2\!\theta_w\, V^{*}_{us}V_{ud} }.
\eea
The short-distance function $Y(x_t)$ describes contribution from Z-penguin and  box diagrams with an internal top quark with QCD corrections. Its expression in NLO can be written as \cite{Misiak:1999yg, Buchalla:1998ba}
\bea
\label{4.3}
Y(x_t) = \eta_Y\cdot \frac{x_t}{8}\left(\frac{4-x_t}{1-x_t}+\frac{3x_t}{(1-x_t)^2}{\rm Ln}\,x_t\right),
\eea
where the factor $\eta_Y$ summarizes the QCD corrections ($\eta_Y = 1.012$). The function $Y_{\rm NL}$  represents the contribution of loop-diagrams 
involving internal charm-quark exchange and is known to NLO \cite{Buchalla:1993wq, Buchalla:1998ba} and recently to NNLO \cite{Gorbahn:2006bm}. The charm contribution is also often denoted by $P_c(Y)$ and is related to $Y_{NL}$ analogous to the relation in Eq~\eqref{3.3}. In the SM, the branching ratio  for the SD part  is written as \cite{Crivellin:2016vjc, Gorbahn:2006bm}
\bea
{\rm BR }(K_L\rightarrow \mu^{+}\mu^{-})_{SM}({\rm SD}) &=& \frac{N_K^2}{2\pi\,\G_{K_L}}\left(\frac{m_\mu}{m_K}\right)^2\sqrt{1-\frac{4\,m_\mu^2}{m_K^2}}\nonumber\\
&& \hskip0.7cm \times\,\, f_K^2m_K^3({\rm Re}\,\D^K_{SM})^2,\nonumber\\
\eea
\begin{figure}[ht!]
 	\centering

 		\hbox{\includegraphics[scale=0.33]{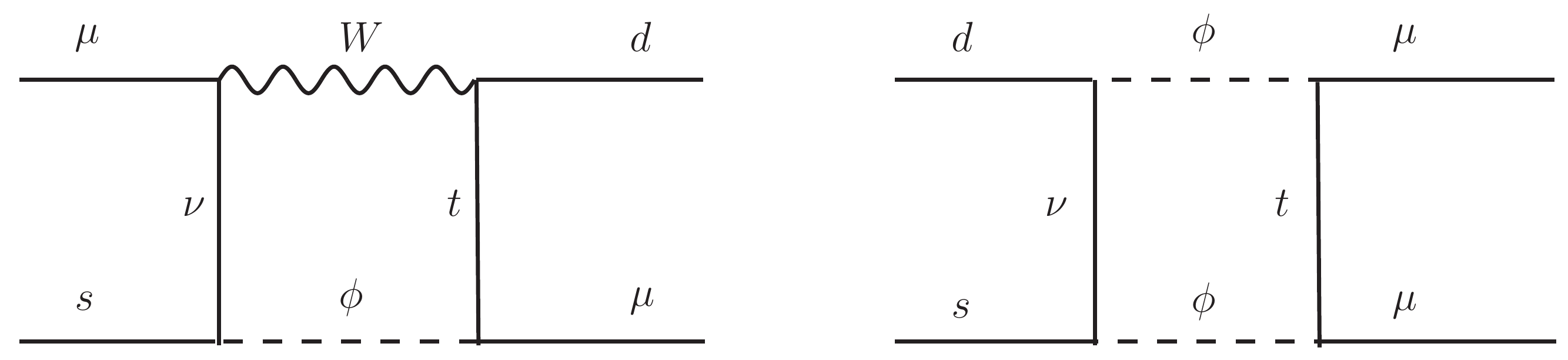}}

 	\caption{ Feynman diagrams relevant for the decay $K_L\rightarrow \mu^{+}\mu^{-}$ induced by the scalar leptoquark $\phi$. 
 	}
 	\label{fig5}
 	 	
 \end{figure} 
 where $N_K = G_F V_{us}^\ast V_{ud}$ and $\G_{K_L}$ is the decay width of $K_L$.  Before proceeding to discuss the constraints on leptoquark couplings from $K_L\rightarrow\mu^{+}\mu^{-}$, we give a description of the ``operator basis" we use in the present and  next sections. The effective Hamiltonian for $K_L\rightarrow\mu^{+}\mu^{-}$ in Eq.~\eqref{4.1} is written in the operator basis of \{$Q_{7V}, \,Q_{7A}$\} following Ref.~\cite{Crivellin:2016vjc}. In what follows, we will switch to the \{$Q^K_{VLL},\,Q^K_{VLR}$\} operator basis. The operators in both  bases are written as
\bea
\label{4.4}
Q_{7V} &=& (\bar{s}\g_\a (1-\g_5) d)(\bar{\mu}\g^\a \mu), \nonumber\\
 Q_{7A} &=& (\bar{s}\g_\a (1-\g_5) d)(\bar{\mu}\g^\a \g_5\mu),
\eea 
and
\bea
\label{4.5}
Q^K_{VLL} &=& (\bar{s}\g_\a L d)(\bar{\mu}\g^\a L\mu), \nonumber\\ 
Q^K_{VLR} &=& (\bar{s}\g_\a L d)(\bar{\mu}\g^\a R\mu).
\eea 
 To change from the basis \{$Q_{7V},\,Q_{7A}$\} to the basis \{$Q^K_{VLL},\,Q^K_{VLR}$\}, the following transformation rules hold:
 \bea
 \label{4.6}
 Q^K_{VLL} = \frac{1}{4}\left(Q_{7V}-Q_{7A}\right),\hskip0.5cm Q^K_{VLR} = \frac{1}{4}\left(Q_{7V}+Q_{7A}\right).\nonumber\\
 \eea
 The scalar leptoquark $\phi$ contributes  to the quark-level transition $\bar{s}\rightarrow \bar{d} \mu^{+}\mu^{-}$ at the leading order through loop diagrams. The Feynman diagrams relevant for $K_L\rightarrow \mu^{+}\mu^{-}$ are shown in Fig~\ref{fig5}. These diagrams are similar to the ones calculated in the case of $b\rightarrow s\mu\mu$ in Ref.~\cite{Bauer:2015knc}. Correcting for the different quark content and coupling, and taking into account the prefactors in the definitions of the effective Hamiltonian for K and B system, we can straightforwardly obtain the result for NP Wilson coefficients of effective operators $Q^K_{VLL}$, $Q^K_{VLR}$  and are given by

\bea
\label{4.7}
C_{VLL}^{K(new)} &=& -\frac{1}{8\pi^2}\frac{\l_t}{\l_u}\frac{m_t^2}{M_\phi^2}\lvert\l_{t\mu}^L\rvert^2 \nonumber\\
&& +\,\, \frac{\sqrt{2}}{64 G_F\pi^2 M_\phi^2}\frac{\xi_{ds}\,\xi^L_{\mu\mu}}{\l_u},\\
C_{VLR}^{K(new)} &=& -\frac{1}{16\pi^2}\frac{\l_t}{\l_u}\frac{m_t^2}{M_\phi^2}\lvert\l_{t\mu}^R\rvert^2\left({\rm Ln}\frac{M_\phi^2}{m_t^2} - f(x_t)\right)\nonumber\\
&& + \,\,\frac{\sqrt{2}}{64 G_F\pi^2 M_\phi^2}\frac{\xi_{ds}\,\xi^R_{\mu\mu}}{\l_u},
\eea
where the function $f(x_t)$ depends on  the top-quark mass and is given in Ref.~\cite{Bauer:2015knc} and we define
\be
\label{4.7.0}
\xi^{L(R)}_{\ell\ell^{'}} = \sum_i\l^{L(R)\ast}_{u_i\ell}\l^{L(R)}_{u_i\ell^{'}}.
\ee

\begin{figure}[ht!]
 	\centering
 	
 	\hbox{\!\!
 		\hbox{\includegraphics[scale=0.35]{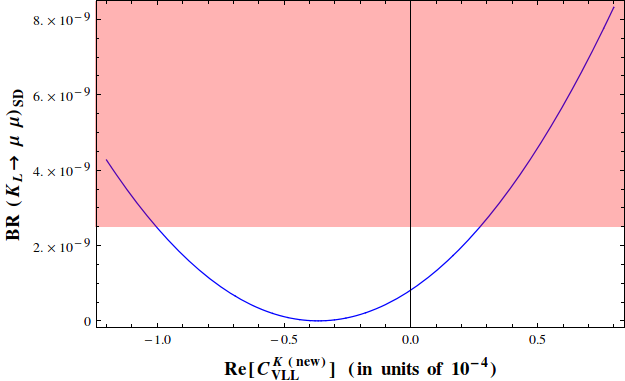}}
 		
 	}
 	
 	\caption{ The dependence of ${\rm BR}(K_L\rightarrow\m^{+}\m^{-})$ on the Wilson coefficient $C_{VLL}^{K(new)}$. We have assumed one-operator dominance as discussed in the text. The red shaded region shows the experimentally disallowed values at $1\sigma$.}
 	\label{fig6}
 	 	
 \end{figure}

 The one advantage we get by the change of basis is that the contribution of  right-handed interaction terms in the Lagrangian (Eq.~\eqref{0.1})  is contained only in $C_{VLR}^{K(new)}$. After accommodating the leptoquark contribution to the SM value, the total SD branching ratio for the decay $K_L\rightarrow\mu^{+}\mu^{-}$ is given by

\bea
\label{4.8}
{\rm BR }(K_L\rightarrow \mu^{+}\mu^{-})_{\rm SD} &=& \frac{N_K^2}{2\pi\,\G_{K_L}}\left(\frac{m_\mu}{m_K}\right)^2\sqrt{1-\frac{4\,m_\mu^2}{m_K^2}}\nonumber\\
&&\times f_K^2m_K^3\l^{10}\left\{\frac{{\rm Re}\l_c}{\l}\frac{\a \,P_c(Y)}{2\pi\sin^2\!\theta_W \l_u}\right.\nonumber\\
&&+\frac{1}{\l^5} \left.\left({\rm Re}\l_t\frac{\a \,Y(x_t)}{2\pi\sin^2\!\theta_W\l_u}\right.\right.\nonumber\\
&&\left.\left.+\frac{1}{4}{\rm Re}(C_{VLR}^{K(new)}-C_{VLL}^{K(new)})\right)\right\}^2,\nonumber\\
\eea

 To simplify further the analysis, we invoke the assumption that except the SM contribution only one of the NP operators contributes dominantly. This assumption helps us in determining the limits on the dominant WC from ${\rm BR}\,(K_L\rightarrow \m^{+}\m^{-})_{\rm SD}$,   and the generalization of this situation to incorporate more than one NP operator contribution is straight forward. Therefore, in what follows, we will ignore the contribution of the right-handed operator in further analysis.
 In Fig~\ref{fig6}, we show the dependence of the SD part of BR($K_L\rightarrow\mu^{+}\mu^{-}$) on  ${\rm Re}\,C_{VLL}^{K(new)}$. Numerically the bound on the WC reads $-1.00\times 10^{-4} < {\rm Re}\,C_{VLL}^{K(new)} < 0.27\times 10^{-4}$. We use the upper bound to  constrain the generation-diagonal leptoquark couplings in the following way. Employing  Eq.~\eqref{4.7}, the upper bound on the WC  can be written in terms of  model parameters as
 \bea 
 \label{4.9a}
\left(-\frac{1}{8\pi^2}\frac{\rm{Re}\,\l_t}{\l_u}\frac{m_t^2}{M_\phi^2}\lvert\l_{t\mu}^L\rvert^2  + \frac{\sqrt{2}}{64 G_F\pi^2 M_\phi^2}\frac{\rm{Re}\,\xi_{ds}}{\l_u}\,\xi^L_{\mu\mu}\right)\nonumber\\
 < 0.27\times 10^{-4}, \eea
 Assuming the worst possible case in which the bound on $\rm{Re}\,\xi_{ds}$ from $K^{+}\rightarrow \pi^{+}\nu\bar{\nu}$ (as obtained in section \ref{sec3}) is saturated, i.e., using $\rm{Re}\,\xi_{ds} =2.2 \times 10^{-4} $ in the above equation, we get 

\bea
\label{4.9}
\sqrt{\lvert\l_{u\mu}^L\rvert^2 + \lvert\l_{c\mu}^L\rvert^2 +\left(1+\frac{2.52}{(\frac{M_\phi}{1000\gev})^2}\right)\lvert\l_{t\mu}^L\rvert^2} < 11.83.\nonumber\\
\eea
We find that constraints from the SD branching ratio of $K_L\rightarrow \m^{+}\m^{-}$ are not severe and large $\sim O(1)$  generation-diagonal leptoquark couplings are allowed. To this end, we must mention that the above bound is in agreement with the constraint obtained in Ref.~\cite{Bauer:2015knc} (see Eq.(17) therein) while explaining the anomaly in $R_K$ in this model.
We also note from Eq.~\eqref{4.9} that the top contribution to $\bar{s}\rightarrow \bar{d}\m^{+} \m^{-}$ for the considered masses  of the leptoquark $(\sim 1 \tev)$ is largely enhanced in contrast to the effects found in the case of $b\rightarrow s\m^{+}\m^{-}$ processes \cite{Bauer:2015knc} where the top contribution is suppressed for the same choice of the leptoquark masses.   

\section{{\boldmath Constraints from LFV decay $K_L \rightarrow \mu^\mp {\lowercase{e}}^\pm$}}
\label{sec6}
In this section, we discuss the effects of the leptoquark $\phi$ on LFV process $K\rightarrow \mu^\mp e^\pm$. Experimentally, there is only an  upper bound on this process: BR$(K_L\rightarrow\mu^\mp e^\pm) < 4.7\times 10^{-12}$ \cite{Agashe:2014kda}. LFV processes are interesting because in the SM they are forbidden. Therefore, any observation of such process immediately indicates towards the presence of NP. The leptoquark $\phi$ can mediate $K_L\rightarrow \mu e$ decay through similar diagrams shown in Fig~\ref{fig5} with one of the $\mu$ lines being replaced with $e$. After integrating out heavy particles, new effective operators relevant for $K_L\rightarrow\mu e$ are generated. The operators are similar to those in  Eq.~\eqref{4.5} but with one of the $\mu$  changed to $e$. The branching ratio in terms of the new Wilson coefficients $C_{VLL}^{\mu e}$ and $C_{VLR}^{\mu e}$ is given by \cite{Crivellin:2016vjc}
\bea
\label{5.1}
{\rm BR}(K_L\rightarrow\mu e) &=& \frac{N_K^2 f_K^2}{64\pi \G_{K_L}}\left(\frac{m_\mu}{m_K}\right)^2\left(1-\frac{m_\mu^2}{m_K^2}\right)^2\nonumber\\
&& \times \left(\lvert C_{VLL}^{\mu e}\rvert^2 + \lvert C_{VLR}^{\mu e}\rvert^2\right),
\eea
Adapting the results of Eq.~\eqref{4.7} to the LFV case, we find

\bea
\label{5.2}
C_{VLL}^{\m e} &=& -\frac{1}{8\pi^2}\frac{\l_t}{\l_u}\frac{m_t^2}{M_\phi^2}(\l^L_{te}\l^{L\ast}_{t\m}) \nonumber\\
&& +\,\, \frac{\sqrt{2}}{64 G_F\pi^2 M_\phi^2}\frac{\xi_{ds}\,\xi^L_{\mu e}}{\l_u},\\
C_{VLR}^{\m e} &=& -\frac{1}{16\pi^2}\frac{\l_t}{\l_u}\frac{m_t^2}{M_\phi^2}(\l^R_{t\m}\l^R_{te})\left({\rm Ln}\frac{M_\phi^2}{m_t^2} - f(x_t)\right)\nonumber\\
&& + \,\,\frac{\sqrt{2}}{64 G_F\pi^2 M_\phi^2}\frac{\xi_{ds}\,\xi^R_{\mu e}}{\l_u}.
\eea
Using the current experimental bound on $K_L\rightarrow\m e$, we get $\left[\lvert C_{VLL}^{\m e}\rvert^2 + \lvert C_{VLR}^{\m e}\rvert^2 \right]^{1/2} < 3.9\times 10^{-6}$. Following the similar analysis as done in section \ref{sec5} for the case of $K_L\rightarrow\m\m$, we obtain the constraints on the leptoquark couplings, 

\begin{eqnarray}
\label{5.3}
\Biggl(\sqrt{(\l_{u\mu}^L \l_{u e}^L) + (\l_{c\mu}^L \l_{c e}^L) +\left(1+\frac{2.52}{(\frac{M_\phi}{1000\gev})^2}\right)(\l_{t\mu}^L \l_{t e}^L)}\Biggr)\nonumber\\
 < 4.49,\nonumber\\
\end{eqnarray}
where the top contribution is again enhanced. For simplicity we assumed the couplings to be real. Here, we would like to mention that the same Wilson coefficients also contribute to other LFV processes such as $K\rightarrow \pi \m e$. However, as pointed out in Ref.~\cite{Crivellin:2016vjc}, the constraints on Wilson coefficients $\left(\lvert C_{VLL}^{\m e}\rvert^2 + \lvert C_{VLR}^{\m e}\rvert^2 \right)^{1/2}$ are about an order of magnitude weaker than the one from $K_L\rightarrow \m^\mp e^\pm$. Therefore, experimental data on $K\rightarrow \pi\m e$ do not improve the constraints obtained in Eq.~\eqref{5.3}.

\section{Results and Discussion}
In light of several anomalies observed in semileptonic B decays, often explained by invoking leptoquark NP models, we have studied a scalar leptoquark model in the context of rare decays of kaons and neutral kaon mixing. The model is interesting because it can provide one of the possible explanations for the observed discrepancies in semileptonic B decays.  We examined the effects of leptoquark contribution to the several kaon processes involving $K^0-\bar{K^0}$ mixing, $K^{+}\rightarrow \pi^{+}\nu\bar{\nu}$, $K_L\rightarrow\pi^0\nu\bar{\nu}$, $K_L\rightarrow \m^{+}\m^{-}$, and LFV decay $K_L\rightarrow \m^{\mp}e^\pm$. Working in the framework of EFT, we have discussed the effective operators generated after integrating out heavy particles and written down the explicit expressions of the corresponding Wilson coefficient in terms of the leptoquark couplings.  Using the present experimental information on these decays, we derived bounds on the couplings relevant for kaon processes. We found that the constraints from $K^0-\bar{K^0}$ on the real and imaginary parts of left-handed coupling $\xi_{ds}$ are $\sim O(10^{-2})$. However, the same set of couplings can also be constrained from ${\rm BR}(K^{+}\rightarrow \pi^{+}\nu\bar{\nu})$, ${\rm BR}(K_L\rightarrow\pi^0\nu\bar{\nu})$, and it was found that constraints from the rare process ${\rm BR}(K^{+}\rightarrow \pi^{+}\nu\bar{\nu})$ are about 2 orders of magnitude more severe than those obtained from the mixing of neutral kaons. In fact, the decay ${\rm BR}(K^{+}\rightarrow \pi^{+}\nu\bar{\nu})$  gives the most stringent constraints on the leptoquark couplings among all the processes studied in this work and therefore is  the most interesting observable to test the NP effects of  the scalar leptoquark in the kaon sector.   Assuming a one-operator dominance scenario, we constrained the NP Wilson coefficient contributing to the rate of $K_L\rightarrow\m^{+}\m^{-}$. We further used the bounds on the NP Wilson coefficient to obtain the constraints on generation-diagonal leptoquark couplings. We found that the present measured value of ${\rm BR}(K_L\rightarrow\m^{+}\m^{-})$ allows generation-diagonal coupling of the leptoquark to be $\sim O(1)$. The  constraint on the combination of generation-diagonal couplings from $K_L \rightarrow \mu^{+}\mu^{-}$ is in agreement with the one obtained in Ref.~\cite{Bauer:2015knc} for explaining experimental data on $R_K$. However, whereas the top contribution to $b\rightarrow s\m^{+}\m^{-}$ is suppressed, we found that in the case of $\bar{s}\rightarrow\bar{d}\m^{+}\m^{-}$ the top contribution is enhanced for the considered range of leptoquark masses. We also did a similar analysis for the case of LFV decay $K_L\rightarrow\m^\mp e^\pm$, which involves generation-diagonal as well as off-diagonal couplings. We found that  present experimental limits on ${\rm BR}\,(K_L\rightarrow\m^\mp e^\pm)$ do not provide very strong constraints, and involved couplings can be as large as $\sim O(1)$. \\

{\bf Acknowledgements:}
The author would like to thank Namit Mahajan, Anjan Joshipura, and Saurabh Rindani for many helpful discussions.  The author would like to thank Monika Blanke for a very useful communication regarding kaon mixing observables. The author  also thanks Abhaya Swain and Chandan Hati for help with the preparation of this manuscript.
 

\begin{thebibliography}{}
 



 \bibitem{Aad:2012tfa} 
   G.~Aad {\it et al.} [ATLAS Collaboration],
   Phys.\ Lett.\ B {\bf 716}, 1 (2012)
   doi:10.1016/j.physletb.2012.08.020
   [arXiv:1207.7214 [hep-ex]].
 
 \bibitem{Chatrchyan:2012xdj} 
   S.~Chatrchyan {\it et al.} [CMS Collaboration],
   Phys.\ Lett.\ B {\bf 716}, 30 (2012)
   doi:10.1016/j.physletb.2012.08.021
   [arXiv:1207.7235 [hep-ex]].

\bibitem{Ricciardi:2015iwa} 
G.~Ricciardi {\it et al.},
Eur.\ Phys.\ J.\ Plus {\bf 130}, no. 10, 209 (2015)
doi:10.1140/epjp/i2015-15209-y
[arXiv:1507.05029 [hep-ph]]; 
J.~Ellis,
arXiv:1604.00333 [hep-ph].

\bibitem{Aaij:2015yra} 
  R.~Aaij {\it et al.} [LHCb Collaboration],
  Phys.\ Rev.\ Lett.\  {\bf 115}, no. 11, 111803 (2015)
  [arXiv:1506.08614 [hep-ex]].
  
\bibitem{Lees:2012xj} 
  J.~P.~Lees {\it et al.} [BaBar Collaboration],
  Phys.\ Rev.\ Lett.\  {\bf 109}, 101802 (2012)
  [arXiv:1205.5442 [hep-ex]];  
  J.~P.~Lees {\it et al.} [BaBar Collaboration],
  Phys.\ Rev.\ D {\bf 88}, no. 7, 072012 (2013)
  [arXiv:1303.0571 [hep-ex]].
  
\bibitem{Huschle:2015rga} 
  M.~Huschle {\it et al.} [Belle Collaboration],
  arXiv:1507.03233 [hep-ex].
  
%


 
 \bibitem{Fajfer:2012vx} 
   S.~Fajfer, J.~F.~Kamenik and I.~Nisandzic,
   Phys.\ Rev.\ D {\bf 85}, 094025 (2012)
   doi:10.1103/PhysRevD.85.094025
   [arXiv:1203.2654 [hep-ph]]; 
  H.~Na {\it et al.} [HPQCD Collaboration],
  Phys.\ Rev.\ D {\bf 92}, no. 5, 054510 (2015) 
   [arXiv:1505.03925 [hep-lat]].
  
  
   \bibitem{Aaij:2014ora} 
     R.~Aaij {\it et al.} [LHCb Collaboration],
     Phys.\ Rev.\ Lett.\  {\bf 113}, 151601 (2014)
     doi:10.1103/PhysRevLett.113.151601
     [arXiv:1406.6482 [hep-ex]].
  
 
 \bibitem{Bobeth:2007dw} 
   C.~Bobeth, G.~Hiller and G.~Piranishvili,
   JHEP {\bf 0712}, 040 (2007)
   doi:10.1088/1126-6708/2007/12/040
   [arXiv:0709.4174 [hep-ph]].
 
 
 
 \bibitem{Hiller:2003js} 
   G.~Hiller and F.~Kruger,
   Phys.\ Rev.\ D {\bf 69}, 074020 (2004)
   doi:10.1103/PhysRevD.69.074020
   [hep-ph/0310219].
 
 
 
 \bibitem{Aaij:2013qta} 
   R.~Aaij {\it et al.} [LHCb Collaboration],
   Phys.\ Rev.\ Lett.\  {\bf 111}, 191801 (2013)
   doi:10.1103/PhysRevLett.111.191801
   [arXiv:1308.1707 [hep-ex]].
   
 \bibitem{Aaij:2015oid} 
   R.~Aaij {\it et al.} [LHCb Collaboration],
   JHEP {\bf 1602}, 104 (2016)
   doi:10.1007/JHEP02(2016)104
   [arXiv:1512.04442 [hep-ex]].
 
 


\bibitem{DescotesGenon:2012zf} 
S.~Descotes-Genon, J.~Matias, M.~Ramon and J.~Virto,
JHEP {\bf 1301}, 048 (2013)
doi:10.1007/JHEP01(2013)048
[arXiv:1207.2753 [hep-ph]].


\bibitem{Matias:2012xw} 
J.~Matias, F.~Mescia, M.~Ramon and J.~Virto,
JHEP {\bf 1204}, 104 (2012)
doi:10.1007/JHEP04(2012)104
[arXiv:1202.4266 [hep-ph]].


\bibitem{Descotes-Genon:2013vna} 
S.~Descotes-Genon, T.~Hurth, J.~Matias and J.~Virto,
JHEP {\bf 1305}, 137 (2013)
doi:10.1007/JHEP05(2013)137
[arXiv:1303.5794 [hep-ph]].

 \bibitem{Aaij:2015esa} 
 R.~Aaij {\it et al.} [LHCb Collaboration],
 JHEP {\bf 1509}, 179 (2015)
 doi:10.1007/JHEP09(2015)179
 [arXiv:1506.08777 [hep-ex]].

\bibitem{Descotes-Genon:2013wba} 
S.~Descotes-Genon, J.~Matias and J.~Virto,
Phys.\ Rev.\ D {\bf 88}, 074002 (2013)
doi:10.1103/PhysRevD.88.074002
[arXiv:1307.5683 [hep-ph]].


 
 \bibitem{Altmannshofer:2015sma} 
 W.~Altmannshofer and D.~M.~Straub,
 arXiv:1503.06199 [hep-ph].
 
\bibitem{Descotes-Genon:2015uva} 
S.~Descotes-Genon, L.~Hofer, J.~Matias and J.~Virto,
JHEP {\bf 1606}, 092 (2016)
doi:10.1007/JHEP06(2016)092
[arXiv:1510.04239 [hep-ph]].



 
 

   
  \bibitem{Celis:2012dk} 
    A.~Celis, M.~Jung, X.~Q.~Li and A.~Pich,
    JHEP {\bf 1301}, 054 (2013)
    [arXiv:1210.8443 [hep-ph]].
    
  \bibitem{Ko:2012sv} 
    P.~Ko, Y.~Omura and C.~Yu,
    JHEP {\bf 1303}, 151 (2013)
    [arXiv:1212.4607 [hep-ph]].
    
  \bibitem{Crivellin:2012ye} 
    A.~Crivellin, C.~Greub and A.~Kokulu,
    Phys.\ Rev.\ D {\bf 86}, 054014 (2012)
    [arXiv:1206.2634 [hep-ph]].

\bibitem{Deshpande:2012rr} 
  N.~G.~Deshpande and A.~Menon,
  JHEP {\bf 1301}, 025 (2013)
  [arXiv:1208.4134 [hep-ph]].
  
 \bibitem{Hati:2015awg} 
   C.~Hati, G.~Kumar and N.~Mahajan,
   JHEP {\bf 1601}, 117 (2016)
   doi:10.1007/JHEP01(2016)117
   [arXiv:1511.03290 [hep-ph]].
  
  
  
  



  
\bibitem{Datta:2012qk} 
  A.~Datta, M.~Duraisamy and D.~Ghosh,
  Phys.\ Rev.\ D {\bf 86}, 034027 (2012)
  [arXiv:1206.3760 [hep-ph]].
  

 \bibitem{Tanaka:2012nw} 
   M.~Tanaka and R.~Watanabe,
   Phys.\ Rev.\ D {\bf 87}, no. 3, 034028 (2013)
   [arXiv:1212.1878 [hep-ph]].

 
\bibitem{Freytsis:2015qca} 
  M.~Freytsis, Z.~Ligeti and J.~T.~Ruderman,
  Phys.\ Rev.\ D {\bf 92}, no. 5, 054018 (2015)
  [arXiv:1506.08896 [hep-ph]].
  
\bibitem{Bhattacharya:2015ida} 
  S.~Bhattacharya, S.~Nandi and S.~K.~Patra,
  arXiv:1509.07259 [hep-ph].
  

  

  
\bibitem{Dorsner:2009cu} 
  I.~Dorsner, S.~Fajfer, J.~F.~Kamenik and N.~Kosnik,
  Phys.\ Lett.\ B {\bf 682}, 67 (2009)
  doi:10.1016/j.physletb.2009.10.087
  [arXiv:0906.5585 [hep-ph]].
 
\bibitem{Sakaki:2013bfa} 
  Y.~Sakaki, M.~Tanaka, A.~Tayduganov and R.~Watanabe,
  Phys.\ Rev.\ D {\bf 88}, no. 9, 094012 (2013)
  [arXiv:1309.0301 [hep-ph]].
  
   
 \bibitem{Fajfer:2012jt} 
   S.~Fajfer, J.~F.~Kamenik, I.~Nisandzic and J.~Zupan,
   Phys.\ Rev.\ Lett.\  {\bf 109}, 161801 (2012)
   [arXiv:1206.1872 [hep-ph]].
  
  
  \bibitem{Dorsner:2013tla} 
    I.~Doršner, S.~Fajfer, N.~Košnik and I.~Nišandžić,
    JHEP {\bf 1311}, 084 (2013)
    doi:10.1007/JHEP11(2013)084
    [arXiv:1306.6493 [hep-ph]].
  


\bibitem{Blake:2015tda} 
  T.~Blake, T.~Gershon and G.~Hiller,
  Ann.\ Rev.\ Nucl.\ Part.\ Sci.\  {\bf 65}, 113 (2015)
  doi:10.1146/annurev-nucl-102014-022231
  [arXiv:1501.03309 [hep-ex]].


\bibitem{Gauld:2013qja} 
  R.~Gauld, F.~Goertz and U.~Haisch,
  JHEP {\bf 1401}, 069 (2014)
  doi:10.1007/JHEP01(2014)069
  [arXiv:1310.1082 [hep-ph]].


\bibitem{Glashow:2014iga} 
  S.~L.~Glashow, D.~Guadagnoli and K.~Lane,
  Phys.\ Rev.\ Lett.\  {\bf 114}, 091801 (2015)
  doi:10.1103/PhysRevLett.114.091801
  [arXiv:1411.0565 [hep-ph]].



\bibitem{Crivellin:2015lwa} 
  A.~Crivellin, G.~D'Ambrosio and J.~Heeck,
  Phys.\ Rev.\ D {\bf 91}, no. 7, 075006 (2015)
  doi:10.1103/PhysRevD.91.075006
  [arXiv:1503.03477 [hep-ph]].


\bibitem{Altmannshofer:2014cfa} 
  W.~Altmannshofer, S.~Gori, M.~Pospelov and I.~Yavin,
  Phys.\ Rev.\ D {\bf 89}, 095033 (2014)
  doi:10.1103/PhysRevD.89.095033
  [arXiv:1403.1269 [hep-ph]].


\bibitem{Buras:2013qja} 
  A.~J.~Buras and J.~Girrbach,
  JHEP {\bf 1312}, 009 (2013)
  doi:10.1007/JHEP12(2013)009
  [arXiv:1309.2466 [hep-ph]].
  
  
\bibitem{Crivellin:2015mga} 
  A.~Crivellin, G.~D'Ambrosio and J.~Heeck,
  Phys.\ Rev.\ Lett.\  {\bf 114}, 151801 (2015)
  doi:10.1103/PhysRevLett.114.151801
  [arXiv:1501.00993 [hep-ph]].

 \bibitem{Chiang:2016qov} 
   C.~W.~Chiang, X.~G.~He and G.~Valencia,
   arXiv:1601.07328 [hep-ph].
   
\bibitem{Boucenna:2016wpr} 
S.~M.~Boucenna, A.~Celis, J.~Fuentes-Martin, A.~Vicente and J.~Virto,
Phys.\ Lett.\ B {\bf 760}, 214 (2016)
doi:10.1016/j.physletb.2016.06.067
[arXiv:1604.03088 [hep-ph]].

    
   
  \bibitem{Gripaios:2014tna} 
    B.~Gripaios, M.~Nardecchia and S.~A.~Renner,
    JHEP {\bf 1505}, 006 (2015)
    doi:10.1007/JHEP05(2015)006
    [arXiv:1412.1791 [hep-ph]].
   
  
  \bibitem{Becirevic:2015asa} 
    D.~Bečirević, S.~Fajfer and N.~Košnik,
    Phys.\ Rev.\ D {\bf 92}, no. 1, 014016 (2015)
    doi:10.1103/PhysRevD.92.014016
    [arXiv:1503.09024 [hep-ph]].
  
   
   \bibitem{Alonso:2015sja} 
     R.~Alonso, B.~Grinstein and J.~M.~Camalich,
     JHEP {\bf 1510}, 184 (2015)
     doi:10.1007/JHEP10(2015)184
     [arXiv:1505.05164 [hep-ph]].
   
   
   \bibitem{Calibbi:2015kma} 
     L.~Calibbi, A.~Crivellin and T.~Ota,
     Phys.\ Rev.\ Lett.\  {\bf 115}, 181801 (2015)
     doi:10.1103/PhysRevLett.115.181801
     [arXiv:1506.02661 [hep-ph]].
   
   
\bibitem{Bauer:2015knc} 
M.~Bauer and M.~Neubert,
Phys.\ Rev.\ Lett.\  {\bf 116}, no. 14, 141802 (2016)
doi:10.1103/PhysRevLett.116.141802
[arXiv:1511.01900 [hep-ph]].


         \bibitem{Hiller:2014yaa} 
           G.~Hiller and M.~Schmaltz,
           Phys.\ Rev.\ D {\bf 90}, 054014 (2014)
           doi:10.1103/PhysRevD.90.054014
           [arXiv:1408.1627 [hep-ph]].
              
   
  
 \bibitem{Barbieri:2015yvd} 
   R.~Barbieri, G.~Isidori, A.~Pattori and F.~Senia,
   Eur.\ Phys.\ J.\ C {\bf 76}, no. 2, 67 (2016)
   doi:10.1140/epjc/s10052-016-3905-3
   [arXiv:1512.01560 [hep-ph]].
   
 
   
   
   \bibitem{Fajfer:2015ycq} 
     S.~Fajfer and N.~Košnik,
     Phys.\ Lett.\ B {\bf 755}, 270 (2016)
     doi:10.1016/j.physletb.2016.02.018
     [arXiv:1511.06024 [hep-ph]].
   
   
   \bibitem{Varzielas:2015iva} 
     I.~de Medeiros Varzielas and G.~Hiller,
     JHEP {\bf 1506}, 072 (2015)
     doi:10.1007/JHEP06(2015)072
     [arXiv:1503.01084 [hep-ph]].
   
    
     
   
   
   \bibitem{Blanke:2013goa} 
     M.~Blanke,
     PoS KAON {\bf 13}, 010 (2013)
     [arXiv:1305.5671 [hep-ph]].
   
   
   \bibitem{Buras:2012ts} 
     A.~J.~Buras and J.~Girrbach,
     Acta Phys.\ Polon.\ B {\bf 43}, 1427 (2012)
     doi:10.5506/APhysPolB.43.1427
     [arXiv:1204.5064 [hep-ph]].
   
   
   
   \bibitem{Buras:2004qb} 
     A.~J.~Buras, T.~Ewerth, S.~Jager and J.~Rosiek,
     Nucl.\ Phys.\ B {\bf 714}, 103 (2005)
     doi:10.1016/j.nuclphysb.2005.02.014
     [hep-ph/0408142].
   
   
   
   \bibitem{Blanke:2009am} 
     M.~Blanke, A.~J.~Buras, B.~Duling, S.~Recksiegel and C.~Tarantino,
     Acta Phys.\ Polon.\ B {\bf 41}, 657 (2010)
     [arXiv:0906.5454 [hep-ph]].
   
   
   \bibitem{Blanke:2008yr} 
     M.~Blanke, A.~J.~Buras, B.~Duling, K.~Gemmler and S.~Gori,
     JHEP {\bf 0903}, 108 (2009)
     doi:10.1088/1126-6708/2009/03/108
     [arXiv:0812.3803 [hep-ph]].
   
   \bibitem{Bauer:2009cf} 
     M.~Bauer, S.~Casagrande, U.~Haisch and M.~Neubert,
     JHEP {\bf 1009}, 017 (2010)
     doi:10.1007/JHEP09(2010)017
     [arXiv:0912.1625 [hep-ph]].
   
   
   \bibitem{Lee:2015qra} 
     C.~J.~Lee and J.~Tandean,
     JHEP {\bf 1508}, 123 (2015)
     doi:10.1007/JHEP08(2015)123
     [arXiv:1505.04692 [hep-ph]].
   
   \bibitem{Buras:2012dp} 
     A.~J.~Buras, F.~De Fazio, J.~Girrbach and M.~V.~Carlucci,
     JHEP {\bf 1302}, 023 (2013)
     doi:10.1007/JHEP02(2013)023
     [arXiv:1211.1237 [hep-ph]].
   
   
   
   \bibitem{Buras:2014yna} 
     A.~J.~Buras, F.~De Fazio and J.~Girrbach-Noe,
     JHEP {\bf 1408}, 039 (2014)
     doi:10.1007/JHEP08(2014)039
     [arXiv:1405.3850 [hep-ph]].
   
   
   \bibitem{Buras:2013ooa} 
   A.~J.~Buras and J.~Girrbach,
   Rept.\ Prog.\ Phys.\  {\bf 77}, 086201 (2014)
   doi:10.1088/0034-4885/77/8/086201
   [arXiv:1306.3775 [hep-ph]].
   
   \bibitem{Buras:1997ij} 
   A.~J.~Buras, A.~Romanino and L.~Silvestrini,
   Nucl.\ Phys.\ B {\bf 520}, 3 (1998)
   doi:10.1016/S0550-3213(98)00169-2
   [hep-ph/9712398].
   
   \bibitem{Queiroz:2014pra} 
     F.~S.~Queiroz, K.~Sinha and A.~Strumia,
     Phys.\ Rev.\ D {\bf 91}, no. 3, 035006 (2015)
     doi:10.1103/PhysRevD.91.035006
     [arXiv:1409.6301 [hep-ph]].
     
\bibitem{Mescia:2012fg} 
F.~Mescia and J.~Virto,
Phys.\ Rev.\ D {\bf 86}, 095004 (2012)
doi:10.1103/PhysRevD.86.095004
[arXiv:1208.0534 [hep-ph]].
     
   
     \bibitem{diphoton excess ref}
        ATLAS and CMS physics results from Run-II, see:  
         {\url{http://indico.cern.ch/event/442432/}}
   
   
    
       \bibitem{Sahoo:2015wya} 
         S.~Sahoo and R.~Mohanta,
         Phys.\ Rev.\ D {\bf 91}, no. 9, 094019 (2015)
         doi:10.1103/PhysRevD.91.094019
         [arXiv:1501.05193 [hep-ph]].
       

         
   \bibitem{Fajfer:2008tm} 
     S.~Fajfer and N.~Kosnik,
     Phys.\ Rev.\ D {\bf 79}, 017502 (2009)
     doi:10.1103/PhysRevD.79.017502
     [arXiv:0810.4858 [hep-ph]].
   
   
   \bibitem{deBoer:2015boa} 
     S.~de Boer and G.~Hiller,
     arXiv:1510.00311 [hep-ph].
   
   
   \bibitem{Fajfer:2015mia} 
     S.~Fajfer and N.~Košnik,
     Eur.\ Phys.\ J.\ C {\bf 75}, no. 12, 567 (2015)
     doi:10.1140/epjc/s10052-015-3801-2
     [arXiv:1510.00965 [hep-ph]].
   
  
   
 	\bibitem{Antonelli:2009ws} 
 	M.~Antonelli {\it et al.},
 	Phys.\ Rept.\  {\bf 494}, 197 (2010)
 	doi:10.1016/j.physrep.2010.05.003
 	[arXiv:0907.5386 [hep-ph]].
 	
 	\bibitem{Buchalla:1995vs} 
 	G.~Buchalla, A.~J.~Buras and M.~E.~Lautenbacher,
 	Rev.\ Mod.\ Phys.\  {\bf 68}, 1125 (1996)
 	doi:10.1103/RevModPhys.68.1125
 	[hep-ph/9512380].
 	
 	\bibitem{Inami:1980fz} 
 	T.~Inami and C.~S.~Lim,
 	Prog.\ Theor.\ Phys.\  {\bf 65}, 297 (1981)
 	[Prog.\ Theor.\ Phys.\  {\bf 65}, 1772 (1981)].
 	doi:10.1143/PTP.65.297
 	
 	\bibitem{Branco:1999fs} 
 	G.~C.~Branco, L.~Lavoura and J.~P.~Silva,
 	Int.\ Ser.\ Monogr.\ Phys.\  {\bf 103}, 1 (1999).


\bibitem{Buras:1990fn} 
A.~J.~Buras, M.~Jamin and P.~H.~Weisz,
Nucl.\ Phys.\ B {\bf 347}, 491 (1990).
doi:10.1016/0550-3213(90)90373-L

\bibitem{Brod:2011ty} 
J.~Brod and M.~Gorbahn,
Phys.\ Rev.\ Lett.\  {\bf 108}, 121801 (2012)
doi:10.1103/PhysRevLett.108.121801
[arXiv:1108.2036 [hep-ph]].

\bibitem{Brod:2010mj} 
J.~Brod and M.~Gorbahn,
Phys.\ Rev.\ D {\bf 82}, 094026 (2010)
doi:10.1103/PhysRevD.82.094026
[arXiv:1007.0684 [hep-ph]].




\bibitem{Bona:2007vi} 
M.~Bona {\it et al.} [UTfit Collaboration],
JHEP {\bf 0803}, 049 (2008)
doi:10.1088/1126-6708/2008/03/049
[arXiv:0707.0636 [hep-ph]]. Updates available on {\url{http://www.utfit.org.}}




\bibitem{Buras:2008nn} 
A.~J.~Buras and D.~Guadagnoli,
Phys.\ Rev.\ D {\bf 78}, 033005 (2008)
doi:10.1103/PhysRevD.78.033005
[arXiv:0805.3887 [hep-ph]].

\bibitem{Buras:2010pza} 
A.~J.~Buras, D.~Guadagnoli and G.~Isidori,
Phys.\ Lett.\ B {\bf 688}, 309 (2010)
doi:10.1016/j.physletb.2010.04.017
[arXiv:1002.3612 [hep-ph]].

  \bibitem{Ligeti:2016qpi} 
    Z.~Ligeti and F.~Sala,
    arXiv:1602.08494 [hep-ph].

\bibitem{Buras:2014maa} 
  A.~J.~Buras, J.~M.~Gérard and W.~A.~Bardeen,
  Eur.\ Phys.\ J.\ C {\bf 74}, 2871 (2014)
  doi:10.1140/epjc/s10052-014-2871-x
  [arXiv:1401.1385 [hep-ph]].
  

  

\bibitem{Buras:2004uu} 
  A.~J.~Buras, F.~Schwab and S.~Uhlig,
  Rev.\ Mod.\ Phys.\  {\bf 80}, 965 (2008)
  doi:10.1103/RevModPhys.80.965
  [hep-ph/0405132].
  
  \bibitem{Mescia:2007kn} 
    F.~Mescia and C.~Smith,
    Phys.\ Rev.\ D {\bf 76}, 034017 (2007)
    doi:10.1103/PhysRevD.76.034017
    [arXiv:0705.2025 [hep-ph]].
    
    \bibitem{Isidori:2005xm} 
      G.~Isidori, F.~Mescia and C.~Smith,
      Nucl.\ Phys.\ B {\bf 718}, 319 (2005)
      doi:10.1016/j.nuclphysb.2005.04.008
      [hep-ph/0503107].
      
      \bibitem{Buchalla:1998ux} 
        G.~Buchalla and G.~Isidori,
        Phys.\ Lett.\ B {\bf 440}, 170 (1998)
        doi:10.1016/S0370-2693(98)01088-0
        [hep-ph/9806501].
        
        \bibitem{Geng:1996kd} 
          C.~Q.~Geng, I.~J.~Hsu and Y.~C.~Lin,
          Phys.\ Rev.\ D {\bf 54}, 877 (1996)
          doi:10.1103/PhysRevD.54.877
          [hep-ph/9604228].
          
          \bibitem{Buchalla:1997kz} 
            G.~Buchalla and A.~J.~Buras,
            Phys.\ Rev.\ D {\bf 57}, 216 (1998)
            doi:10.1103/PhysRevD.57.216
            [hep-ph/9707243].
            
            \bibitem{Buchalla:1993wq} 
              G.~Buchalla and A.~J.~Buras,
              Nucl.\ Phys.\ B {\bf 412}, 106 (1994)
              doi:10.1016/0550-3213(94)90496-0
              [hep-ph/9308272].
            
\bibitem{Buchalla:1993bv} 
G.~Buchalla and A.~J.~Buras,
Nucl.\ Phys.\ B {\bf 400}, 225 (1993).
doi:10.1016/0550-3213(93)90405-E

\bibitem{Misiak:1999yg} 
M.~Misiak and J.~Urban,
Phys.\ Lett.\ B {\bf 451}, 161 (1999)
doi:10.1016/S0370-2693(99)00150-1
[hep-ph/9901278].

\bibitem{Buchalla:1998ba} 
G.~Buchalla and A.~J.~Buras,
Nucl.\ Phys.\ B {\bf 548}, 309 (1999)
doi:10.1016/S0550-3213(99)00149-2
[hep-ph/9901288].

\bibitem{Brod:2010hi} 
J.~Brod, M.~Gorbahn and E.~Stamou,
Phys.\ Rev.\ D {\bf 83}, 034030 (2011)
doi:10.1103/PhysRevD.83.034030
[arXiv:1009.0947 [hep-ph]].



\bibitem{Buras:2005gr} 
A.~J.~Buras, M.~Gorbahn, U.~Haisch and U.~Nierste,
Phys.\ Rev.\ Lett.\  {\bf 95}, 261805 (2005)
doi:10.1103/PhysRevLett.95.261805
[hep-ph/0508165].

\bibitem{Buras:2006gb} 
A.~J.~Buras, M.~Gorbahn, U.~Haisch and U.~Nierste,
JHEP {\bf 0611}, 002 (2006)
[JHEP {\bf 1211}, 167 (2012)]
doi:10.1007/JHEP11(2012)167, 10.1088/1126-6708/2006/11/002
[hep-ph/0603079].

\bibitem{Buras:1998raa} 
A.~J.~Buras,
hep-ph/9806471.

\bibitem{Buras:2015qea} 
A.~J.~Buras, D.~Buttazzo, J.~Girrbach-Noe and R.~Knegjens,
JHEP {\bf 1511}, 033 (2015)
doi:10.1007/JHEP11(2015)033
[arXiv:1503.02693 [hep-ph]].

\bibitem{Agashe:2014kda} 
K.~A.~Olive {\it et al.} [Particle Data Group Collaboration],
Chin.\ Phys.\ C {\bf 38}, 090001 (2014).
doi:10.1088/1674-1137/38/9/090001

\bibitem{Ritchie:1993ua} 
  J.~L.~Ritchie and S.~G.~Wojcicki,
  Rev.\ Mod.\ Phys.\  {\bf 65}, 1149 (1993).
  doi:10.1103/RevModPhys.65.1149
  
  
  
  \bibitem{Cirigliano:2011ny} 
    V.~Cirigliano, G.~Ecker, H.~Neufeld, A.~Pich and J.~Portoles,
    Rev.\ Mod.\ Phys.\  {\bf 84}, 399 (2012)
    doi:10.1103/RevModPhys.84.399
    [arXiv:1107.6001 [hep-ph]].
  
  \bibitem{Isidori:2003ts} 
    G.~Isidori and R.~Unterdorfer,
    JHEP {\bf 0401}, 009 (2004)
    doi:10.1088/1126-6708/2004/01/009
    [hep-ph/0311084].
    
    \bibitem{Crivellin:2016vjc} 
      A.~Crivellin, G.~D'Ambrosio, M.~Hoferichter and L.~C.~Tunstall,
      arXiv:1601.00970 [hep-ph].

  
  \bibitem{Gorbahn:2006bm} 
    M.~Gorbahn and U.~Haisch,
    Phys.\ Rev.\ Lett.\  {\bf 97}, 122002 (2006)
    doi:10.1103/PhysRevLett.97.122002
    [hep-ph/0605203].
    
   
     
\end{thebibliography}
\end{document}